\documentclass[showpacs,amsmath,amssymb,aps,showkeys,floatfix,prd,a4paper]{revtex4}

\usepackage[dvips]{graphicx}
\usepackage{dcolumn}
\usepackage{bm}
\usepackage{epsfig}
\usepackage{amsfonts}
\usepackage{amssymb,amscd}
\def\reg{{I\!\!R}}

\def\lsim{\raise0.3ex\hbox{$<$\kern-0.75em\raise-1.1ex\hbox{$\sim$}}}
\def\gsim{\raise0.3ex\hbox{$>$\kern-0.75em\raise-1.1ex\hbox{$\sim$}}}

\def\pom{{I\!\!P}}

\def\beq{\begin{equation}}
\def\eeq{\end{equation}}
\def\bea{\begin{eqnarray}}
\def\eea{\end{eqnarray}}
\def\bq{\begin{quote}}
\def\eq{\end{quote}}

\newcommand{\rb}{\mbox{\boldmath $b$}}

\def\gappeq{\mathrel{\rlap {\raise.5ex\hbox{$>$}}
{\lower.5ex\hbox{$\sim$}}}}

\def\lappeq{\mathrel{\rlap{\raise.5ex\hbox{$<$}}
{\lower.5ex\hbox{$\sim$}}}}

\def\Toprel#1\over#2{\mathrel{\mathop{#2}\limits^{#1}}}

\def\pom{{I\!\!P}}

\begin{document}


\title{Exclusive vector meson photoproduction in fixed - target collisions at the LHC}

\author{V.~P. Gon\c{c}alves and M. M. Jaime}
\affiliation{High and Medium Energy Group, \\
Instituto de F\'{\i}sica e Matem\'atica, Universidade Federal de Pelotas\\
Caixa Postal 354, CEP 96010-900, Pelotas, RS, Brazil}

\date{\today}

\begin{abstract}
The exclusive $\rho$, $\omega$ and $J/\Psi$ photoproduction in fixed - target  collisions at the LHC  is investigated.    We estimate, for the first time, the rapidity and transverse momentum distributions of the vector meson photoproduction in $p He$, $p Ar$, $Pb He$ and $Pb Ar$ fixed - target collisions at the LHC using the STARlight Monte Carlo and present our results for the total cross sections. Predictions for the kinematical range probed by the LHCb detector are also presented. Our results indicate that the experimental analysis of this process in fixed - target collisions at the LHC is feasible. Such future analysis will  probe the QCD dynamics in a kinematical range complementary to that studied in the  collider mode. 
 
\end{abstract}

\maketitle

\section{Introduction}
\label{intro}

The Large Hadron Collider (LHC) at CERN  started high energy collisions nine years ago. During this period a large amount of data have been collected considering $pp$ collisions at $\sqrt{s}$ = 0.9, 2.76, 7, 8 and 13 TeV, $pPb$ collisions at $\sqrt{s} =$ 5 and 8.2 TeV  as well as $PbPb$ collisions at $\sqrt{s}$ = 2.76 and 5 TeV.  Currently, there is a great expectation that LHC will discover  new physics beyond the Standard Model, such as supersymmetry or extra dimensions. However, we should remember that one of the main contributions of the LHC is that it probes a new kinematical regime at high energy, where several questions related to the description of the Quantum
Chromodynamics (QCD) remain without  satisfactory answers. In particular, the study of the photon - induced interactions in hadronic collisions at the LHC \cite{upc,review_forward} is expected to constrain  the nuclear and nucleon gluon distributions \cite{bert} and the description of the QCD dynamics at high energies \cite{vicmag}, where a hadron becomes a dense system and  the nonlinear effects inherent to  the QCD dynamics may become visible \cite{hdqcd}. During the last years, the study of these interactions in $pp/pA/AA$ collisions \cite{upc} became a reality \cite{cdf,star,phenix,alice,alice2,alice3,lhcb,lhcb2,lhcb3,lhcbconf} and new data associated to the Run 2 of the LHC are expected to be released soon.

A complementary kinematical range can be studied in fixed - target collisions at the LHC. Such alternative, proposed originally in Ref. \cite{after}, is expected to probe for example the nucleon and nuclear matter in the  domain of high Feynman $x_F$, the transverse spin asymmetries in the Drell - Yan and Quarkonium production  as well as the Quark Gluon Plasma formation in the energy and density range between the SPS and RHIC experiments \cite{after2}. The basic idea of the AFTER@LHC experiment is to develop a fixed - target programme using the proton and heavy ions beams of the LHC, extracted by a bent crystal, to collide with a fixed proton or nuclear target, reaching high luminosities when a target with a high density is considered. The typical energies that are expected to be reached in this experiment are $\sqrt{s} \approx 110$ GeV for $pA$ collisions and $\sqrt{s} \approx 70$ GeV for $PbA$ collisions. Very recently, the study of fixed - target collisions at the LHC became a reality by the injection of noble gases ($He, \, Ne, \, Ar$) in the LHC beam pipe by the LHCb Collaboration \cite{lhcfixed} using the System for Measuring Overlap with Gas (SMOG) device \cite{smog}. The typical fixed - target $pA$ and $PbA$ configurations that already have been performed were $pAr$ collisions at $\sqrt{s} = 69$ GeV, $pNe$ at $\sqrt{s} = 87$ GeV, $pHe/\, pNe/\, pAr$ at $\sqrt{s	} = 110$ GeV, $Pb Ne$ at $\sqrt{s} = 54$ GeV and  $Pb Ar$ at $\sqrt{s} = 69$ GeV.  The associated experimental results are expected to improve our understanding of the nuclear effects present in $pA$ collisions \cite{lhcbD} and, in the particular case of $pHe$ collisions, to shed light on the antiproton production (See e.g. Ref. \cite{antiproton}).  

The study of photon - induced interactions also is expected to be possible  in fixed - target collisions \cite{after}. As demonstrated in Refs. \cite{lands_dilepton,vicodderon2}, the analysis of the exclusive dilepton and $\eta_c$ photoproduction in ultraperipheral collisions at AFTER@LHC can be useful to probe the inner hadronic structure and the existence of the Odderon, which is one the main open questions of the strong interactions theory \cite{review_odderon}.  Assuming $\sqrt{s} = 115/\,72/\,72$ GeV for $pp/Pbp/PbPb$ collisions,  the typical values for the maximum photon - hadron and photon - photon center - of - mass energies are $\sqrt{s_{\gamma h}} \lesssim 44/\, 12/\, 9$ GeV and $\sqrt{s_{\gamma \gamma}} \lesssim 17/\,2.0/\,1.0$ GeV, respectively.  Therefore, the fixed - target collisions allows to probe the photon - induced interactions in a  limited energy range, dominated by low - energy interactions. Such analysis can be considered as complementary to those performed  in the collider mode, where the maximum energies can reach values of ${\cal{O}}(TeV)$  and the cross sections receive contributions of low and high energies $\gamma \gamma$ and $\gamma h$ interactions \cite{upc}. 

In this paper we will study, for the first time, the exclusive vector meson photoproduction in fixed - target collisions. In particular, we will estimate the rapidity and transverse momentum distributions for the exclusive $\rho$, $\omega$ and $J/\Psi$ photoproduction in $pA$ and $PbA$ collisions at the energies and configurations considered by the LHCb experiment. Our goal in this exploratory study is to provide a first estimate of the total cross sections and associated distributions that can be measured at the LHC taking into account of some realistic experimental requirements on the final states. In our analysis we will use the STARlight Monte Carlo \cite{starlight}  to treat the vector meson photoproduction and its decay. Recent results demonstrate that predictions of this Monte Carlo is able to successfully describe the main aspects of the photon - induced interactions at the LHC (See e.g. Refs. \cite{alice,alice2,lhcb,lhcb2,lhcb3}). We postpone for a future publication a more detailed discussion about the description of the vector meson production in the soft and/or low energy regimes probed in fixed target collisions (See e.g. Ref. \cite{vicmagvector}).  Before to present our results in Section \ref{res},  we will present in the next Section a brief review of photon - induced interactions and its description in the STARlight MC. Finally, in Section \ref{conc} we will summarize our main conclusions and results.

\begin{figure}[t]
\includegraphics[scale=0.3]{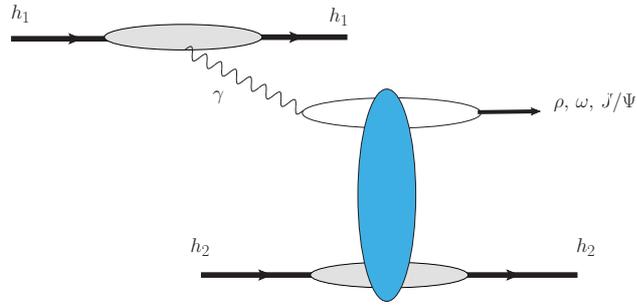}  
\caption{Vector meson photoproduction in a hadronic $h_1 h_2$ collision. The vertical blue ellipse represents the strong interaction between the vector meson and the hadronic target.  }
\label{diagram}
\end{figure}

\section{Formalism}
\label{form}

 The basic idea in  photon-induced processes is that an ultra relativistic charged hadron (proton or nucleus) 
gives rise to strong electromagnetic fields, such that the photon stemming from the electromagnetic field of one of the two  hadrons can either
interact with one photon of the other hadron (photon - photon process) or can interact directly with the other hadron (photon - hadron process) \cite{upc}. 
In these processes the total cross section  can be factorized in terms of the equivalent flux of photons of the incident hadrons and the photon-photon 
or photon-target  cross section. In what follows we will focus in the exclusive vector meson production in photon - hadron interactions. In this process, represented in Fig. \ref{diagram}, the  topology of the final state  will be characterized by two
empty regions  in pseudo-rapidity, called rapidity gaps, separating the intact hadrons from the vector meson. These events also will be characterized by a low hadronic multiplicity and one vector meson with  small transverse momentum. Such characteristics  can be used, in principle, to separate the photon - induced processes from the inelastic one \cite{review_forward}.  Our focus will be in ultraperipheral collisions (UPCs), characterized by large impact parameters ($b > R_{h_1} + R_{h_2}$), in which  the photon -- 
induced interactions become dominant.  In this case, the cross section for the exclusive vector meson photoproduction in photon - induced interactions can be expressed by,
\begin{widetext}
\begin{equation}
   \sigma(h_1 + h_2 \rightarrow h_1 \otimes V \otimes h_2;\,s) =  \int d\omega \,\, n_{h_1}(\omega) \, \sigma_{\gamma h_2 \rightarrow V \otimes h_2}\left(W_{\gamma h_2}  \right) + \int d \omega \,\, n_{h_2}(\omega)
   \, \sigma_{\gamma h_1 \rightarrow V \otimes h_1}\left(W_{\gamma h_1}  \right)\,  \; , 
\label{eq:sigma_pp}
\end{equation}
\end{widetext}
where $\sqrt{s}$ is center-of-mass energy for the $h_1 h_2$ collision ($h_i$ = p,A), $\otimes$ represents the presence of a rapidity gap in the final state, $\omega$ is the energy of the photon emitted by the hadron and $n_h$ is the equivalent photon flux of the hadron $h$ integrated over the impact parameter. Moreover, $\sigma_{\gamma h \rightarrow V \otimes h}$ describes the vector meson production in photon - hadron interactions. In the STARlight MC, the photon spectrum is calculated as follows \cite{klein,starlight}
\begin{eqnarray}
n(\omega) = \int \mbox{d}^{2} {\mathbf b} \, P_{NH} ({\mathbf b}) \,   N\left(\omega,{\mathbf b}\right) \,\,, 
\end{eqnarray}
where $P_{NH} ({\mathbf b})$ is the probability of not having a hadronic interaction at impact parameter ${\mathbf b}$  and the number of photons per unit area, per unit energy, derived assuming a point-like form factor, is given by 
\begin{equation}
N(\omega,{\mathbf b}) = \frac{Z^{2}\alpha_{em}}{\pi^2} \frac{\omega}{\gamma^{2}}
\left[K_1^2\,({\zeta}) + \frac{1}{\gamma^{2}} \, K_0^2({\zeta}) \right]\,
\label{fluxo}
\end{equation}
where $\zeta \equiv \omega b/\gamma$ and $K_0(\zeta)$ and  $K_1(\zeta)$ are the
modified Bessel functions.  Different models for $P_{NH} ({\mathbf b})$ are assumed for $pp$, $pA$ and $AA$ collisions (For details see Ref. \cite{starlight}). Additionally,  the description of the cross section for the $\gamma h \rightarrow V \otimes h$ process depends if the target is a proton or a nuclei. In the proton case, the vector meson photoproduction is described by a parametrization inspired in the Regge theory given by
\begin{eqnarray}
\sigma_{\gamma p \rightarrow V \otimes p} = \sigma_{\pom} \times W_{\gamma p}^{\epsilon} + \sigma_{\reg} \times W_{\gamma p}^{\eta} \,\,,
\label{sig_gamp}
\end{eqnarray}
where the first term is associated to a Pomeron exchange, dominant at high energies, and the second one to the Reggeon exchange, which describes the behavior of the cross section at low energies. For the $\rho$ and $\omega$ production, both terms contribute, with $\epsilon = 0.22$, and $\eta$ being negative, in the range $-1.2 \ge \eta \ge -1.9$. Therefore, the cross section for light mesons rises slowly with increasing $W_{\gamma p}$. On the other hand, the $J/\Psi$ production is assumed to be described only by the Pomeron contribution, with $\epsilon = 0.65$, and the cross section is supplemented by a factor that accounts for its behavior for energies near  the  threshold of production. The free parameters on the parametrization are fitted using the HERA data \cite{hera}.  On the other hand, in the nuclear case, the STARlight allows to estimate the vector meson photoproduction in incoherent ($\gamma A \rightarrow V A^*$) and coherent ($\gamma A \rightarrow V A$) photon - nucleus interactions. In what follows we will present our results for the coherent production, where the cross section is determined using a classical Glauber approach \cite{klein,glauber}. In this case the coherent cross section is given by
\begin{widetext}
\begin{eqnarray}
\sigma(h_1 h_2 \rightarrow h_1 \otimes V \otimes h_2) =  \int_0^{\infty} d\omega \,n_{h_1}(\omega) \, \int_{t_{min}}^{\infty} dt \, \frac{d\sigma(\gamma h_2 \rightarrow V h_2)}{dt}|_{t=0} \,|F(t)|^2 +  [h_1  \leftrightarrow    h_2]\,\,
\label{star1}
\end{eqnarray}
\end{widetext}
where $F(t)$ is the nuclear form factor and $t_{min} = (M_V^2/4 \omega \gamma)^2$. For heavy ions, the form factor is assumed to be the convolution of a hard sphere potential with a Yukawa potential of range 0.7 fm. Such form factor implies the presence of diffractive minima when $t$ is a multiple of $(\pi/R_A)^2$ if nuclear shadowing and saturation effects are negligible.   The differential cross section for a photon - nucleus interaction is determined using the optical theorem and the generalized vector dominance model (GVDM) \cite{sakurai}
\begin{eqnarray}
\frac{d\sigma(\gamma A \rightarrow V A)}{dt}|_{t=0} = \frac{\alpha \sigma^2_{tot}(VA)}{4 f_v^2} \,\,,
\label{gvdm}
\end{eqnarray}
where $f_v$ is the vector  meson - photon coupling and the total cross section for the vector meson - nucleus interactions is found using the classical Glauber calculation
\begin{eqnarray}
\sigma_{tot}(VA) = \int d^2\rb \{ 1 - \exp[-\sigma_{tot}(Vp)T_{AA}(\rb)]\}
\label{glauber}
\end{eqnarray}
with $\sigma_{tot}(Vp)$ being determined by  $\sigma(\gamma p \rightarrow Vp)$ [See Eq. (9) in Ref. \cite{klein}] and $T_{AA}$ is the overlap function at a given impact parameter $\rb$. The Eq. (\ref{glauber}) is denoted classical  since it disregards 
the fact that at high energies the wave package that propagates through the nucleus can be different from the projective wave function and nondiagonal $V - V^{\prime}$ terms can contribute for the total $VA$ cross section. Such corrections were estimated originally in Ref. \cite{frank02}, which demonstrated that the quantum calculation implies higher cross sections. Recent results \cite{alice3} indicate that the classical Glauber gives a better description of the experimental data for the $\rho$ photoproduction in ultraperipheral collisions. The discrepancy between the quantum calculation and data is explained as being associated to the presence of nuclear shadowing effects \cite{frank16}. Consequently, the classical calculation can still be considered a good first approximation to estimate the magnitude of the vector meson  photoproduction cross  sections.

Finally, it is important to emphasize that the STARlight assumes a dipole form factor for a proton target. Such form factor implies a $t$ spectrum that decreases with increasing $t$, without the presence of diffractive dips. Such behavior differs from the results obtained in Refs. \cite{armesto,diego} using phenomenological models based on saturation physics, which predict that dips should be present in the spectrum.

\begin{widetext}
\begin{center}
\begin{table}[t]
\begin{tabular}{|c|c|c|c|c|c|c|}
\hline 
Final State &  p-Ar & p-He & Pb-Ar & Pb-He\tabularnewline
\hline 
\hline 
$\rho^{0}\rightarrow\pi^{+}\pi^{-}$ & 318.60 (16.50) $\mu b$ & 6.97 (1.09) $\mu b$ & 42.50 (24.50) mb & 5.60 (2.44) mb\tabularnewline
\hline 
$\omega\rightarrow\pi^{+}\pi^{-}$  & 1160.12 (30.71) nb & 21.86 (2.29)  nb  & 76.32 (46.21) $\mu b$ & 12.81 (5.35) $\mu b$\tabularnewline
\hline 
$J/\psi\rightarrow\mu^{+}\mu^{-}$ &  3.88 (0.14) nb & 118.41 (14.29) pb & 88.67 (39.68) nb & 13.31 (7.15) nb\tabularnewline
\hline 
\hline 
\end{tabular}
\caption{Total cross sections for the exclusive $\rho$, $\omega$ and $J/\Psi$ photoproduction in fixed - target collisions at the LHC considering $pA$ ($Pb A$) collisions at $\sqrt{s} = 110 \,(69)$ GeV. The predictions obtained assuming the LHCb requirements are presented in parenthesis.}
\label{table:XSeC}
\end{table}
\end{center}
\end{widetext}

\begin{figure}[t]
\begin{tabular}{cc}
\includegraphics[scale=0.35]{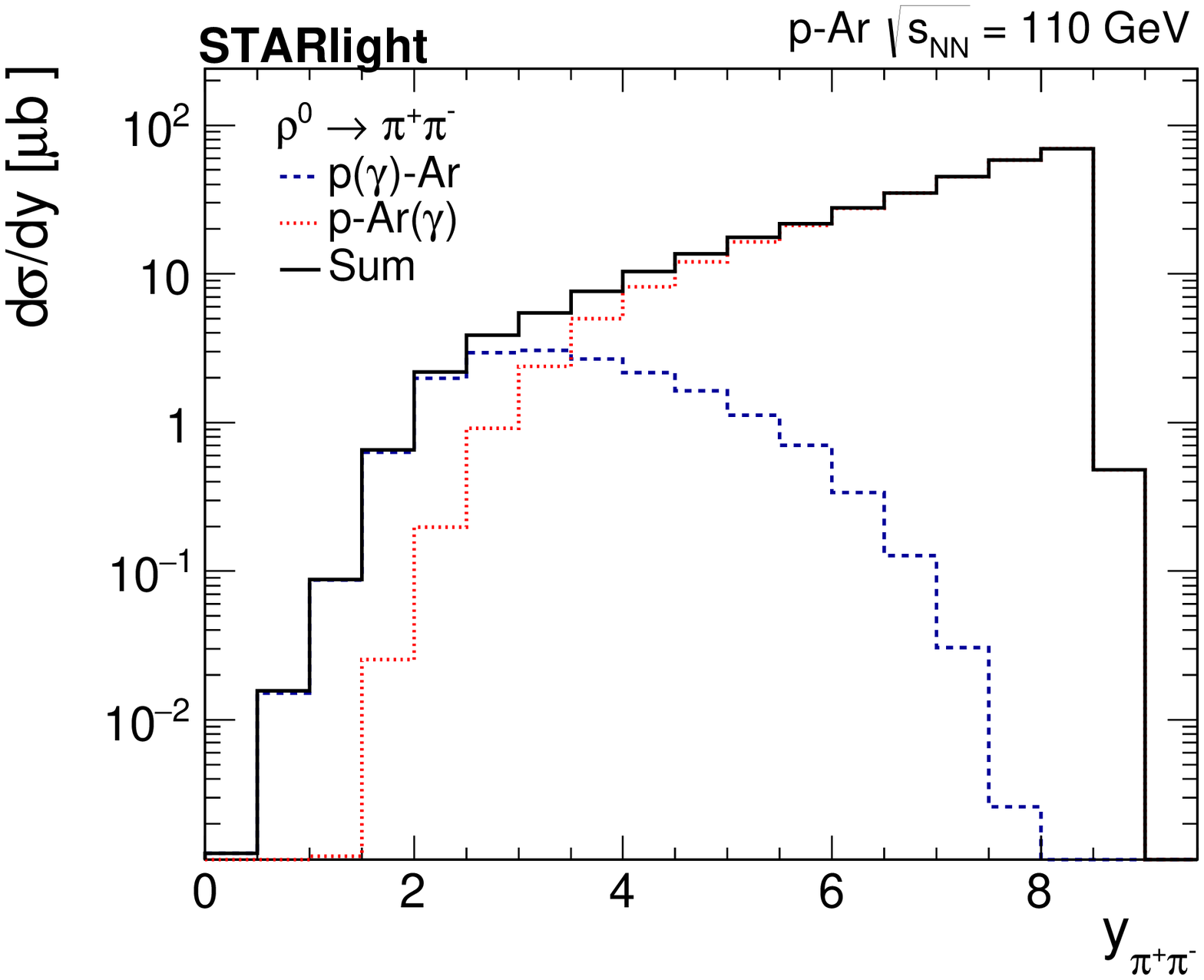} & 
\includegraphics[scale=0.35]{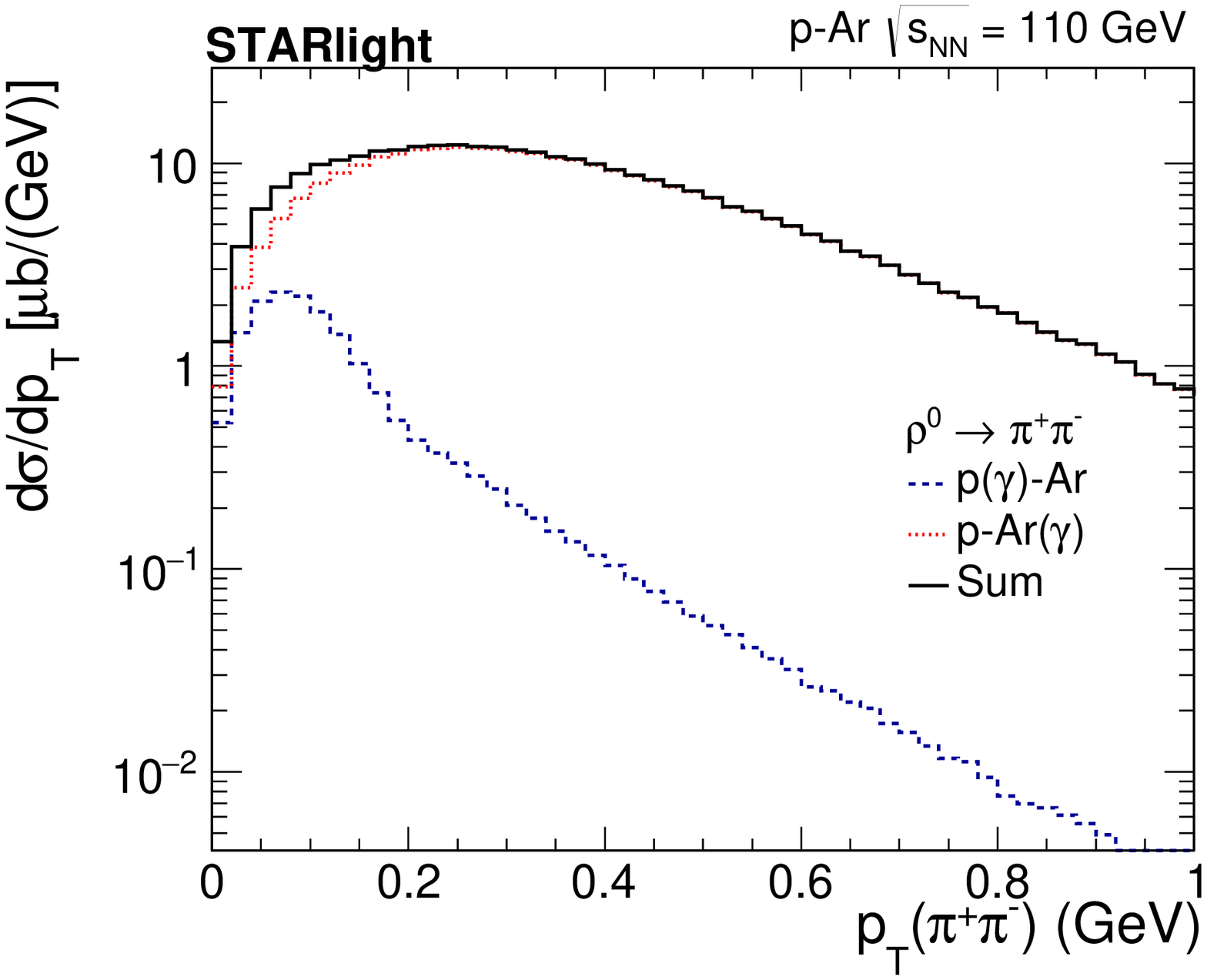}
\end{tabular}
\caption{Rapidity (left panel) and transverse momentum (right panel) distributions for the exclusive $\rho$ photoproduction in $pAr$ collisions at $\sqrt{s} = 110$ GeV. The predictions associated to $\gamma p$ and $\gamma Ar$ interactions are presented separately, as well as the sum of both contributions. The $\gamma$ in parenthesis indicates the particle that is the source of the photons.}
\label{fig:rho}
\end{figure}

\section{Results}
\label{res}

In what follows we will present our estimates for the exclusive $\rho$, $\omega$ and $J/\Psi$ photoproduction in fixed - target $pA$ and $Pb A$ collisions at $\sqrt{s} = 110$ GeV and 69 GeV, respectively, assuming $A = He, Ar$.   The masses and widths are the standard Particle Data Group values. In the case of $\rho$ and $\omega$ production we will present our predictions taking into account the decay of these vector mesons  into  $\pi^+ \pi^-$ pairs. For $J/\Psi$ production, we present our predictions considering its decay in a $\mu^+ \mu^-$ pair. 
Finally, we consider the full LHC kinematical range as well as the kinematical range probed by the LHCb detector. In the latter case, we select the events in which the  
$\pi^+ \pi^-$ and $\mu^+ \mu^-$ pairs are produced in the pseudorapidity range $2 \le \eta \le 5$ with $p_T \ge 0.2$ GeV. 

Our predictions for the total cross sections are presented in Table \ref{table:XSeC}. In our analysis we consider that both initial state particles can be the source of the photons that generate the photon - hadron interactions. As expected from previous studies of the exclusive vector meson photoproduction (See e.g. Refs. \cite{vicmag,brunoall}), the cross sections decrease with the increasing of the mass of the vector meson, being substantially larger for the $\rho$ production.  Moreover, the $PbA$ collisions are enhanced by the $Z^2$ factor present in the nuclear photon flux. We have that our predictions for the light meson production are  approximately three orders of magnitude smaller than those expected in $pPb$ and $PbPb$ collisions at the Run2 of the LHC in the collider mode \cite{brunoall}. In the $J/\Psi$ case, the predictions are smaller by four orders of magnitude, which is directly associated to the fact that $\sigma(\gamma h \rightarrow J/\Psi \otimes h)$ has a steeper energy dependence  than the cross section for the light meson production ($\epsilon_{J/\Psi} = 0.65 \gg \epsilon_{\rho,\omega} = 0.22$) and that in the collider mode  a larger range of values of $W_{\gamma h}$ contributes for the hadronic cross section. In comparison to the predictions for RHIC energies \cite{vicmag}, we have that our results are smaller by approximately two orders of magnitude. However, as already emphasized above, the fixed - target collisions at the LHC are expected to reach high luminosities [${\cal{O}}(100 \, nb^{-1}$) per year], which implies that approximately 
$16\times 10^9$ ($34000$) events per year will be associated to a $\rho$ ($J/\Psi$) produced in an exclusive photon - hadron interaction. As a consequence, the experimental analysis of this process in fixed - target collisions at the LHC is, in principle, feasible. Another important aspect is that if we assume the LHCb requirements for the selection of a exclusive process, the impact on the total cross sections for $PbA$ collisions is small (See Table \ref{table:XSeC}), which implies that the LHCb detector ideal for the study of this process. In contrast, for $pA$ collisions, the LHCb requirements has a large impact on the predictions. In order to understand this result, in Fig. \ref{fig:rho} (left panel) we present the rapidity distribution for the exclusive $\rho$ photoproduction in $pAr$ collisions at $\sqrt{s} = 110$ GeV. The contributions associated to $\gamma p$ and $\gamma Ar$ interactions are presented separately, as well the sum of both. We have that at small rapidities ($y_{\pi^+ \pi^-} \le 3.0$) the distribution is dominated by $\gamma Ar$ interactions, with the photon coming from the proton. This contribution decreases at large rapidities. On the other hand, for $y_{\pi^+ \pi^-} > 3.0$, the $\gamma p$ interactions, present when the photon is emitted by the nucleus,  become dominant and the maximum occurs at very forward rapidities, beyond those probed by the LHCb detector. As a consequence, the LHCb requirements has a large impact on the prediction of the total cross section in $pAr$ collisions. Moreover, as the $\gamma p$ contribution increases with $Z^2$, the impact of these requirements is larger in $pAr$ than in $pHe$ collisions, as verified in Table \ref{table:XSeC}. For completeness, in Fig. \ref{fig:rho} (right panel) we also present our predictions for the transverse momentum distribution. We can see that this distribution  is also dominated by $\gamma p$ interactions. Such behaviour is expected, since the $t$ - behaviour of the $\gamma Ar$ interactions is defined by the nuclear form factor, while in the case of $\gamma p$ the behavior is determined by the nucleon form factor. Due to the difference of values between these quantities, it is expected a narrower transverse momentum distribution in $\gamma Ar$ than in $\gamma p$ interactions. We have verified that similar conclusions are derived analysing the exclusive $\omega$ and $J/\Psi$ photoproduction in $pA$ collisions. In what follows we will only present the sum of $\gamma p$ and $\gamma A$ contributions.

\begin{widetext}

\begin{figure}[t]
\begin{tabular}{ccc}
\includegraphics[scale=0.3]{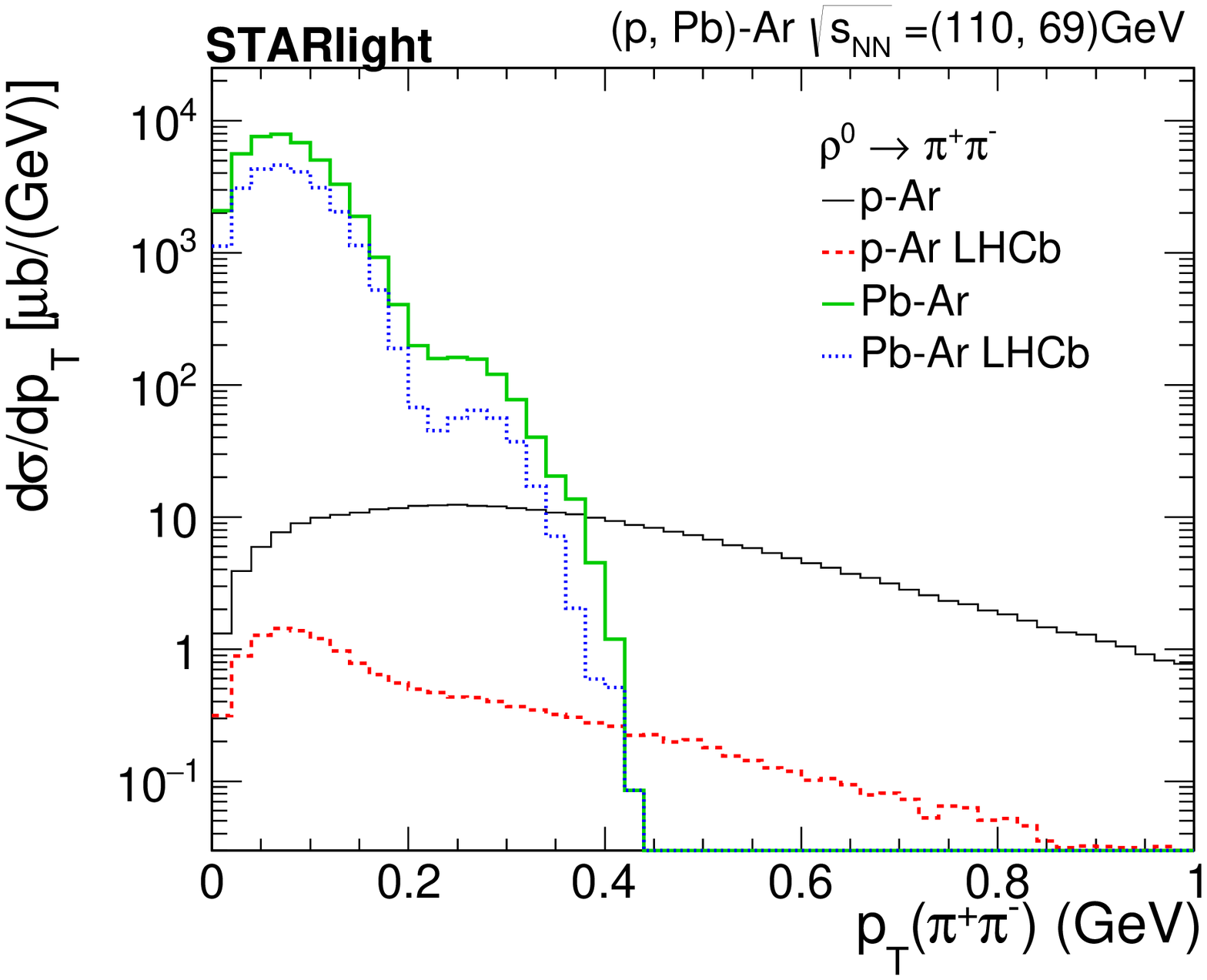} & 
\includegraphics[scale=0.3]{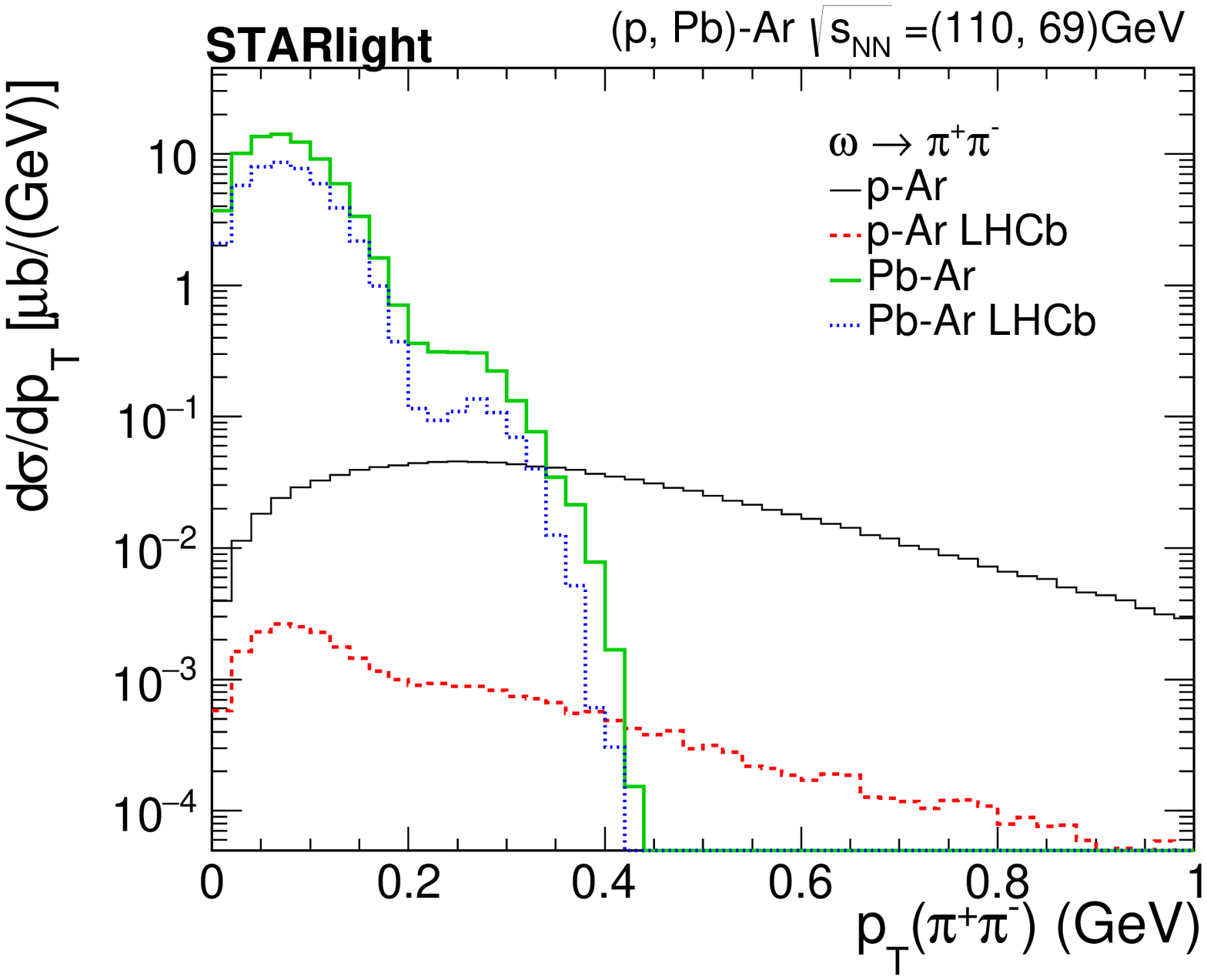} & 
\includegraphics[scale=0.3]{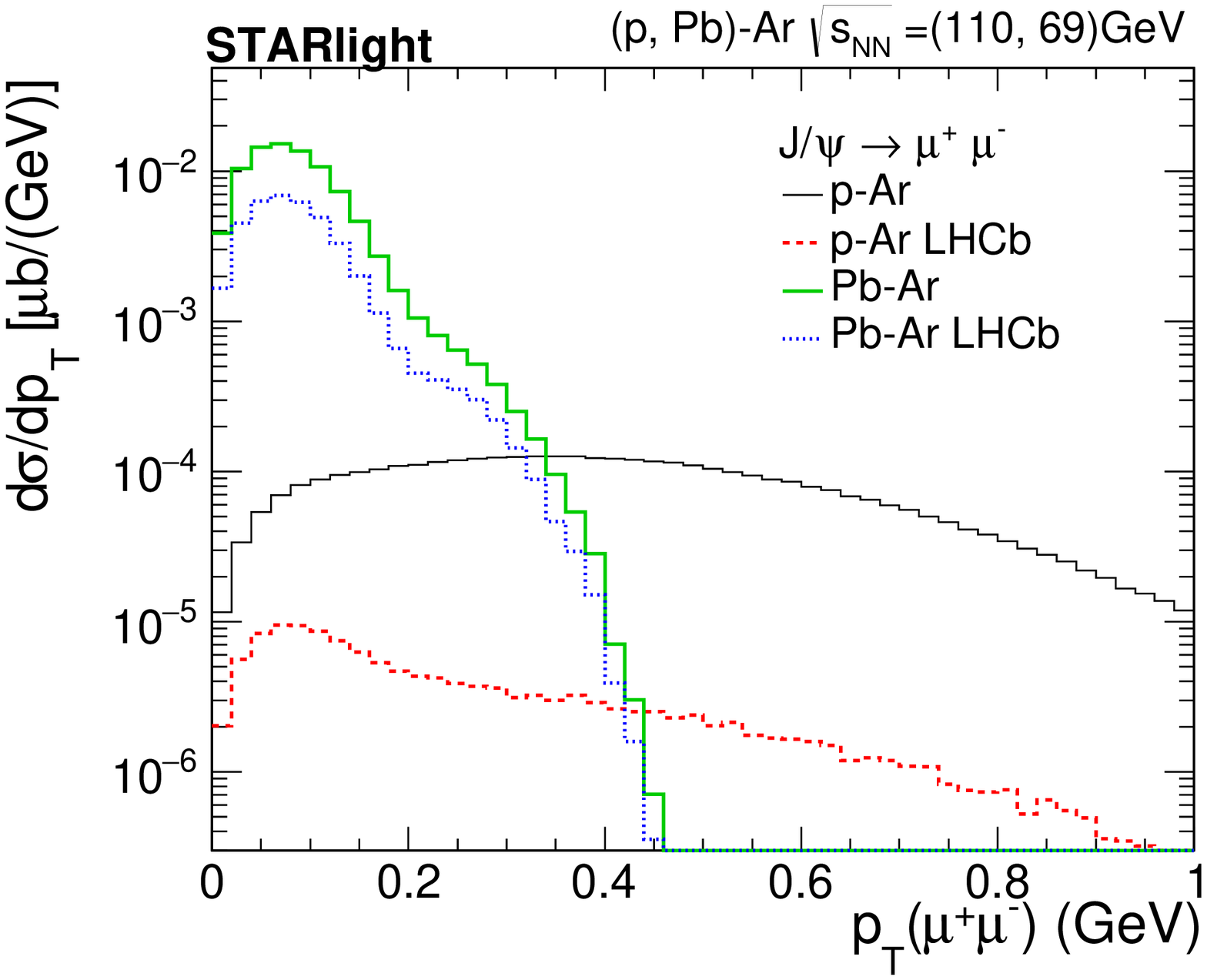}
\end{tabular}
\caption{Transverse momentum distributions for the exclusive vector meson photoproduction in $pAr$ and $PbAr$ collisions at $\sqrt{s} = 110$ and 69 GeV, respectively. The predictions obtained assuming the LHCb requirements are also presented for comparison.  }
\label{fig:pt}
\end{figure} 

\end{widetext}

In Fig. \ref{fig:pt} we present our predictions for the transverse momentum distributions associated to the exclusive $\rho$, $\omega$ and $J/\Psi$ photoproduction in $pAr$ and $PbAr$ collisions at $\sqrt{s} = 110$ and 69 GeV, respectively. 
For comparison, we also present the predictions obtained assuming the LHCb requirements. 
The meson $p_T$ spectrum is determined by the sum of the photon momentum with the exchanged momentum in the interaction between the vector meson and the target. The photon momentum is defined by the equivalent photon approximation, while the exchanged one is determined in the case of coherent interactions by the form factor of the target. As pointed out in Ref. \cite{klein_int}, in the case of symmetric collisions ($h_1 = h_2$), the overall $p_T$ spectrum also is affected by interference from the two production sources. Such effect will be not present in our case, since we are only considering asymmetric collisions ($h_1 \neq h_2$). As both the photon and scattering transverse momenta are small, we expect that meson $p_T$ spectrum will be dominated by small values of transverse momentum, being strongly suppressed at large $p_T^2$. This behaviour is observed in Fig. \ref{fig:pt}.
 Considering the predictions for $pAr$ collisions, as demonstrated before, the $p_T$ distribution is determined by $\gamma p$ interactions, with the photon coming from the nucleus. As a consequence, the associated predictions for the production of the different vector mesons are distinct from those obtained in the case of $PbAr$ collisions, which are determined by $\gamma A$ interactions. In the $pAr$ case, we predict wider distributions.  In the case of $PbAr$ collisions, the $p_T$ - behavior of the distributions is determined by the nuclear form factor $F(t)$ [See Eq. (\ref{star1})], which implies a faster decreasing at large transverse momenta in comparison to the proton case. We have that the main impact of the LHCb requirements is the modification of the normalization of the distributions. 

\begin{widetext}

\begin{figure}[t]
\begin{tabular}{ccc}
\includegraphics[scale=0.3]{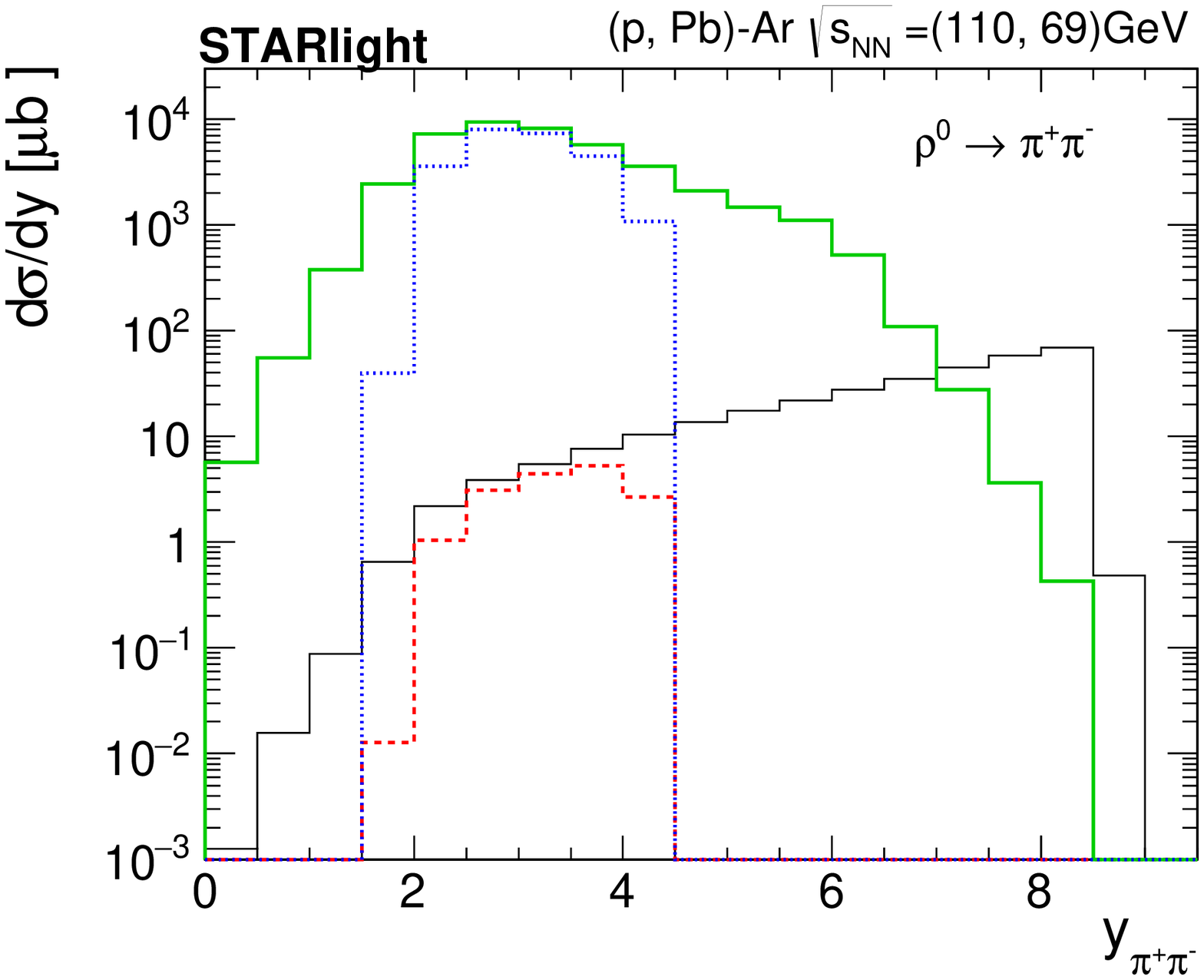} & \includegraphics[scale=0.3]{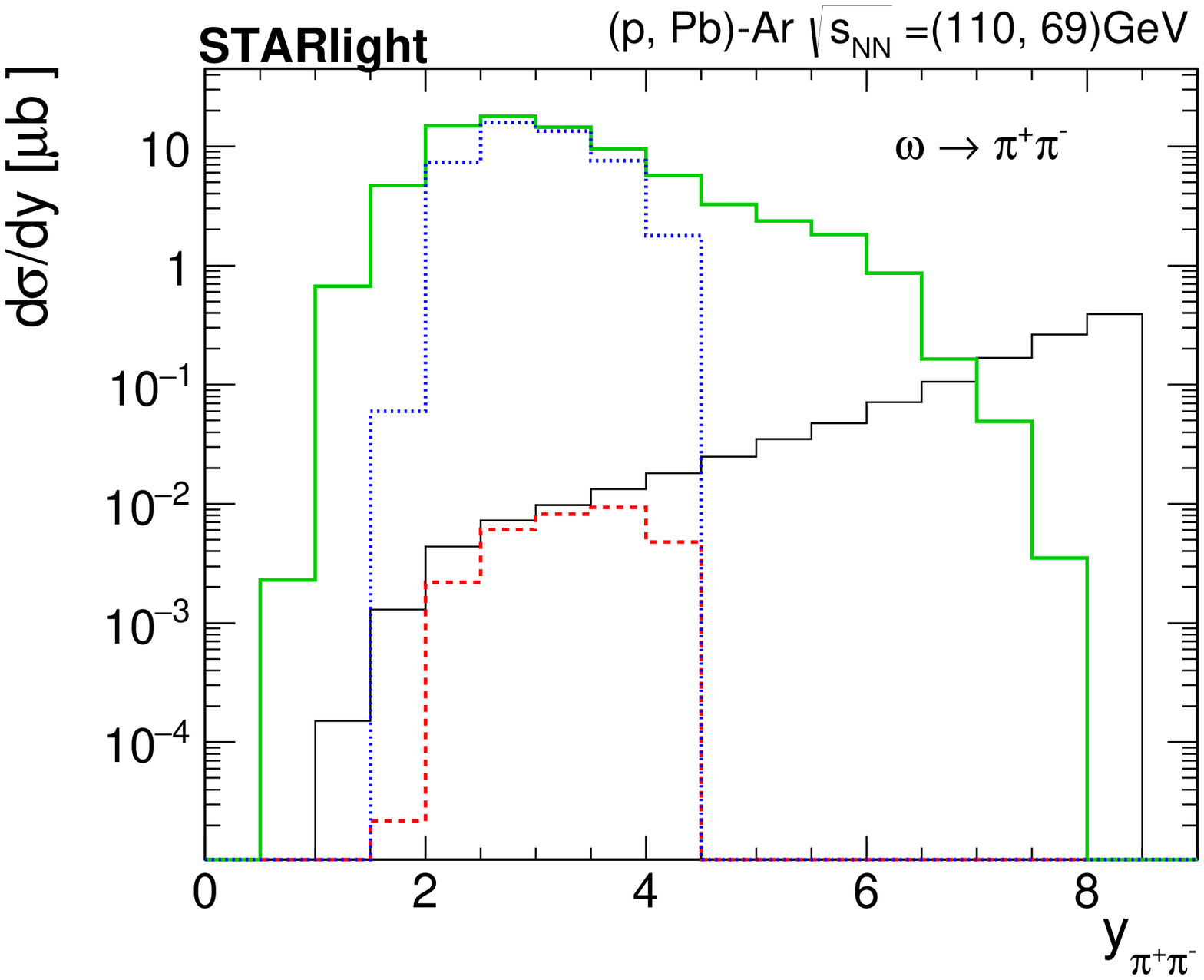} &\includegraphics[scale=0.3]{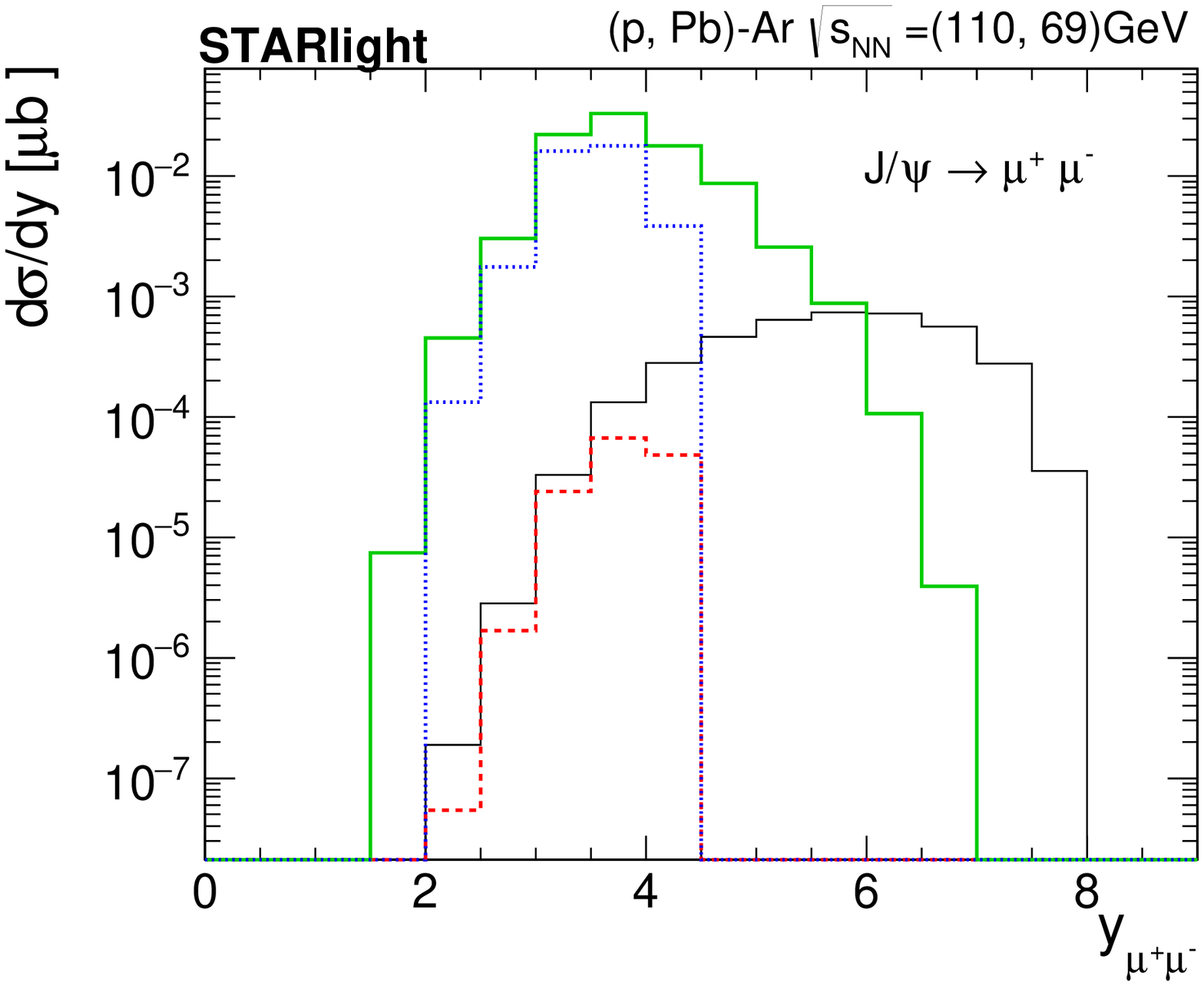}
\end{tabular}
\caption{Rapidity distributions for the exclusive vector meson photoproduction in $pAr$ and $PbAr$ collisions at $\sqrt{s} = 110$ and 69 GeV, respectively. The predictions obtained assuming the LHCb requirements are also presented for comparison.   }
\label{fig:rap}
\end{figure}

\end{widetext}

In Fig. \ref{fig:rap} we present our predictions for the rapidity distributions. As we are considering the collision of non - identical hadrons, the magnitude of the photon fluxes associated to the two incident hadrons is different, which implies asymmetric rapidity distributions. Moreover, distinctly from the predictions for the collider mode, where the maximum of the distributions occur at central rapidities ($y \approx 0$), we have that in fixed - target collisions, the maximum is shifted for forward rapidities.  In particular, it occurs in the kinematical range probed by the LHCb detector in the case of $PbAr$ collisions. As discussed before, this result explains the small impact of the LHCb requirements on the predictions for the total cross sections for $PbAr$ collisions presented in Table \ref{table:XSeC}. 
In the case of light meson production, the rapidity distribution is determined by the Reggeon and Pomeron contributions [See Eq. (\ref{sig_gamp})], with the cross section being determined by the Reggeon one at low energies near the threshold of production. Such energies are probed at small ($y \le 2$) and large ($y \ge 7$) rapidities.  Our results indicate that the Reggeon contribution will be strongly reduced in the kinematical range probed by the LHCb detector. The absence of a reggeonic term  and the higher threshold in the $J/\Psi$ case imply a narrower rapidity distribution.  On the other hand, in the case of $pAr$ collisions, the maximum of the distributions occurs at very forward rapidities, beyond those probed by the LHCb. Although the LHCb requirements have a large impact on the predictions, in particular at larger nuclei, the values presented in Table \ref{table:XSeC} indicate that the analysis of the exclusive vector meson photoproduction in $pA$ collisions is still  feasible. Finally, our results indicate a faster increasing of the rapidity distribution for the $J/\Psi$ production with increasing of $y$ in the region $y\le 4$ in comparison to predicted for the other mesons. Such behaviour is directly associated to the steeper energy behaviour of the $\gamma p \rightarrow J/\Psi p$ cross section, which can be directly probed by the LHCb detector.

\section{Summary}
\label{conc}
During the last years, the experimental results from Tevatron, RHIC and LHC have demonstrated that the study of  hadronic physics using photon - induced interactions in $pp/pA/AA$ colliders is feasible and provide important  information about the QCD dynamics and vector meson production. In this paper, we have 
complemented previous analysis of the exclusive vector meson photoproduction by considering, for the first time, the possibility of study this process in fixed - target collisions at the LHC. Our analysis has been motivated by  the proposition of the AFTER@LHC experiment and by the study of beam - gas interactions, recently performed by the LHCb detector. We have estimated the total cross sections, rapidity and transverse momentum distributions for the exclusive $\rho$, $\omega$ and $J/\Psi$ photoproduction in $pA$ and $PbA$ collisions at $\sqrt{s} = 110$ and $69$ GeV using the STARlight Monte Carlo, which allowed to take into account some typical LHCb requirements for the selection of exclusive events. Our results indicate that the experimental analysis of this process is, in principle, feasible. If performed, such analysis will probe the vector meson production at low energies, which will allows to improve the description of this process in a kinematical regime unexplorated by previous fixed - target experiments and current colliders.   

\section*{Acknowledgements}
VPG would like to thanks M. S. Rangel, J. G. Contreras and S. R. Klein for useful discussions.
This work was  partially financed by the Brazilian funding
agencies CNPq, CAPES,  FAPERGS and  INCT-FNA (process number 
464898/2014-5).

{\it Note added in the proof --} Recently we became aware that the study of exclusive $J/\Psi$ photoproduction in fixed target collisions as a probe of the generalized parton distribution $E_g$ has been briefly discussed in Ref. \cite{lansfixed}.



\end{document}